\documentclass[%
 aip,
 sd,%
 amsmath,amssymb,
 reprint,%
]{revtex4-1}

\usepackage{graphicx}
\usepackage{dcolumn}
\usepackage{bm}

\begin{document}

\preprint{AIP/123-QED}

\title{Spin current evolution in the separated spin-up and spin-down quantum hydrodynamics}
\author{Mariya Iv. Trukhanova}
\email{mar-tiv@yandex.ru}

\affiliation{Faculty of physics, Lomonosov Moscow State University, Moscow, Russian Federation.}

\date{\today}%

\begin{abstract}
We have developed the quantum hydrodynamic model that describes particles with spin-up and with spin-down in separate. We have derived the equation of the spin current evolution as a part of the set of the quantum hydrodynamics (QHD) equations that treat particles with different projection of spin on the preferable direction as two different species. We have studied orthogonal propagation of waves in the external magnetic field and determined the contribution of quantum corrections due to the Bohm potential and to magnetization energy of particles with different projections of spin in the spin current wave dispersion. We have analyzed the limits of weak and strong magnetic fields.
\end{abstract}

 \maketitle

        \section{\label{sec:level1} Introduction}

Spin current is a very important concept of physics. For instance it is useful in spintronics and in the physics of quantum plasma of spinning particles. Spin current generation, detection and manipulation are of the particular interest for magnetoelectronic devices research. Establishment of methods for injecting of spin currents and manipulating spin information has been reported in Refs. \cite{4} - \cite{8}.  The demonstration of spin current injection and detection at room temperature using the geometries and interfaces between ferromagnetic electrodes in combination with a tunnel barrier and nonmagnetic metal was implemented by Jadema et al. \cite{5} - \cite{52}. The main idea of these experiments was that the spin-polarized current in spin-coupled systems leads to charge imbalance. The direct electronic measurement of the Hall-effect-induced spin current in a lateral geometry has been demonstrated in Ref. \cite{4} and the induced voltage in the conductor that results from the conversion of the injected spin current into charge imbalance owing to the spin-orbit coupling was measured.

The magnetic domain wall displacement by spin-polarized current had been confirmed using the magnetic wires \cite{6}, \cite{7}. The real-space observation of current-driven magnetic domain wall (DW) displacement using a well-defined single DW in a micro-fabricated magnetic wire was demonstrated in Ref. \cite{8}. The nature of the influence of spin diffusion in current-induced domain wall motion was theoretically investigated in Ref. \cite{10}. The effect of the conduction electrons' spins on spatial and temporal magnetization dynamics of a ferromagnetic wire was theoretically researched recently \cite{9}, \cite{11} in the time-dependent semiclassical transport theory.

Spin-polarized currents are studied using the phenomenon of spin transfer torque in which the angular momentum of a spin polarized electrical current entering a ferromagnet is transferred to the magnetization \cite{111} - \cite{113} and the spin pumping phenomenon that is the transfer of the spin angular momentum from magnetization precession motion to conduction-electron spin \cite{114}. Spin-polarized currents are used in spin-diode structures \cite{115}, \cite{116} and in observation of the spin Peltier effect \cite{117}, \cite{118}. It was shown that the spin Peltier effect requires independent heat transport in spin-up and spin-down channels \cite{119}.

 The collective motion of spin - spin wave -  can carry non-equilibrium spin currents and  transfer a signal in some magnetic insulators \cite{120} and it has been shown that a spin-wave spin current may persist at much greater distances.

The theoretical description of spin currents is of substantial interest. The definition of the spin-current and the equation of the spin-current evolution had been derived earlier\cite{100}. The spin current equation derived from the many-particle microscopic Schrodinger equation was used to study the dispersion of collective excitations in three dimensional samples of magnetized dielectrics. It had been shown that the dynamics of spin current leads to formation of a new type of collective excitations in magnetized dielectrics, which were called spin-current waves. The existence of pure and persistent spin current without an accompanying charge current in a semiconducting mesoscopic device with a spin-orbit interaction and the definition of the spin current in the presence of spin-orbit interaction had been investigated in Ref. \cite{00}.

The satisfactory description of spin transport in solids was presented in \cite{1} - \cite{3}.
The phenomenological spin continuity equation that describes spin accumulation in solids and contains novel expression for the spin current was presented in Ref. \cite{2}.

On the other hand systems of spin-up and spin-down particles are in the focus of many studies. A quantum hydrodynamics model of charged spin-1/2 particles was developed in Refs. \cite{12}, \cite{133}. The set of equations, which separately describes spin-up electrons and spin-down electrons, was derived:  the continuity equation (particle number evolution equation), the Euler equation (momentum balance equation) and  the Bloch equation (magnetic moment balance equation). Application of the separated spin evolution quantum hydrodynamics to two-dimensional electron gas in plane samples and nanotubes placed in external magnetic fields have revealed the spin-electron acoustic wave in the electron gas \cite{133}.    In this paper we present further application of separated spin evolution QHD.

In this paper we  consider spin-up and spin-down particles  as two different species.
We present our derivation of equations for the spin current evolution, which are a part of the set of QHD equations that describe in separate the evolution of spin-up and spin-down particles moving in an external magnetic field. Corresponding spin current evolution equations were directly derived from the Pauli equation.  This
derivation can be performed in context  of
quantum hydrodynamics method which was developed in Refs. \cite{20} - \cite{22}. The first significant results in the direction of development of the quantum hydrodynamics of degenerate electrons considering two different spin
states (spin-up and spin-down) as two different species
of particles were presented in \cite{12}, \cite{133} and \cite{23}. In the Ref. \cite{23} authors had presented the QHD model of spin-1/2 quantum plasmas. The QHD model contains equations for evolution of the spin density projections, but does not contain the spin current evolution equations. The spin-current appears in the equations for evolution of the spin density projections and  gives contribution in the force field
in the  equation of motion. Thus it is important to have the equations
of spin-current dynamical evolution.      Separated spin evolution QHD and new spin-current dynamical evolution equation   show itself as an usefultool for research of spin transport in magnetic nanostructures. Using the developed model we have analyzed the contribution of the spin polarization to the spin current wave dispersion.   We apply the developed model to the magnetized dielectrics research.

\section{\label{sec:level2} Governing equations}
In Ref.  \cite{12}  it was shown that the evolution of charged spin-1/2 particles with spin-up and with spin-down is governed by the equations for time evolution of the {\em concentration of particles $n_{\uparrow}(\textbf{r},t)$ and $n_{\downarrow}(\textbf{r},t)$},
which are proportional to the probability density

\begin{equation} \label{n1}
\partial_{t}n_{\uparrow}+\nabla(n_{\uparrow}\cdot\textbf{v}_{\uparrow})=\frac{\gamma}{\hbar}(B_yS_x-B_xS_y),
\end{equation} and

\begin{equation} \label{n2}
\partial_{t}n_{\downarrow}+\nabla(n_{\downarrow}\cdot\textbf{v}_{\downarrow})=\frac{\gamma}{\hbar}(B_xS_y-B_yS_x),
\end{equation}
and for time evolution of the particle {\em currents for
each projection of spin}
$$mn_{\uparrow}(\partial_{t}+\textbf{v}_{\uparrow}\cdot \nabla)\textbf{v}_{\uparrow}+\nabla p_{\uparrow}-\frac{\hbar^2}{4m}n_{\uparrow}\nabla\Biggl(\frac{\triangle n_{\uparrow}}{n_{\uparrow}}-\frac{(\nabla n_{\uparrow})^2}{2n^2_{\uparrow}}\Biggr)
$$

$$=qn_{\uparrow}(\textbf{E}+\frac{1}{c}\textbf{v}_{\uparrow}\times\textbf{B})+\gamma n_{\uparrow}\nabla B_z+\frac{\gamma}{2}(S_x\nabla B_x+S_y\nabla B_y)$$

\begin{equation} \label{j1}\qquad\qquad\qquad\qquad+\frac{m\gamma}{\hbar}(\textbf{J}^xB_y-\textbf{J}^yB_x),\end{equation}

   {\em and}

$$mn_{\downarrow}(\partial_{t}+\textbf{v}_{\downarrow}\cdot\nabla)\textbf{v}_{\downarrow}+\nabla p_{\downarrow}-\frac{\hbar^2}{4m}n_{\downarrow}\nabla\Biggl(\frac{\triangle n_{\downarrow}}{n_{\downarrow}}-\frac{(\nabla n_{\downarrow})^2}{2n^2_{\downarrow}}\Biggr)$$

$$=qn_{\downarrow}(\textbf{E}+\frac{1}{c}\textbf{v}_{\downarrow}\times\textbf{B})-\gamma n_{\downarrow}\nabla B_z+\frac{\gamma}{2}(S_x\nabla B_x+S_y\nabla B_y)$$

\begin{equation} 
\label{j2}\qquad\qquad\qquad\qquad+\frac{m\gamma}{\hbar}(\textbf{J}^yB_x-\textbf{J}^xB_y),\end{equation} where $q$ stands for the charge of the particles, for electrons ($q=-e$), $m$ - is the mass of the particles, $\gamma$ - is the gyromagnetic ratio, for electrons $\gamma=-g|e|\hbar/2m_ec$ and $g\simeq1.00116$ - is the g-factor. In the equations (\ref{n1}) - (\ref{j2}) we have the following physical quantities:  $n_{\uparrow,\downarrow}(\textbf{r},t)$ - are the concentrations of particles  in the point $\textbf{r}$ and  $\textbf{v}_{\uparrow,\downarrow}(\textbf{r},t)$ -  are the velocity fields of particles baring spin-up (spin-down), $S_h$ - are the projections of the spin density vector and $p_{\uparrow,\downarrow}(\textbf{r},t)$ - are the thermal pressures.    The second and third terms at the left-hand side of Euler equations (\ref{j1}) and (\ref{j2}) represent gradients of the thermal pressure and of the Bohm potential. The second and third  terms at the right-hand side describe the effect of the $z$-projection, $x$- and $y$ - projections of magnetic field on spin densities of particles and characterize the parts of force field which represents the influence  of the magnetic field on magnetic moments.  The last group of terms is related to nonconservation of particle number with different spin-projection. This nonconservation provides a mechanism for the change of the momentum density in the extra force fields \cite{12}.

It is important to generalize the spin density evolution equations in the separated spin evolution quantum hydrodynamics. The spin density  evolution $S_x$ and $S_y$ is given by  \cite{12}

\begin{equation} \label{s1}
\partial_{t}S_x+\nabla\textbf{J}_x=\frac{2\gamma}{\hbar}(B_zS_y-B_y(n_{\uparrow}-n_{\downarrow})),
\end{equation}
{\em and}

\begin{equation} \label{s2}
\partial_{t}S_y+\nabla\textbf{J}_y=\frac{2\gamma}{\hbar}(B_x(n_{\uparrow}-n_{\downarrow})-B_zS_x),
\end{equation}
where $\textbf{J}_x$ and $\textbf{J}_y$ - are spin currents. The first and second terms at right sides of equations  (\ref{s1}) and (\ref{s2}) represent the action of torque exerted by a magnetic field on a magnetic texture.  The spin density projection on the $"z"$ direction presents the difference between concentrations of particles with different projection of spin $S_z=n_{\uparrow}-n_{\downarrow}$ and the spin-current $\textbf{J}_z=\textbf{j}_{\uparrow}-\textbf{j}_{\downarrow}$ is the difference between the particle currents of particles with different projection of spin $\textbf{j}_{\uparrow,\downarrow}$. To make our study of spin density dynamics more detailed we present the direct derivation of equation for the spin current evolution from the many-particle Schrodinger equation and determine its influence on the spin current wave dispersion.

\subsection{Spin current evolution}

Next, in this section we are going to represent the spin-current evolution equations for degenerate particles considering spin-up and spin-down states as two different species.
Particles were studied as single fluid.
We use the Pauli equation for a single particle in an external electromagnetic field \cite{12} and self-consistent field approximation \cite{22}.   We start from the Pauli  equation         for spin-up $\psi_{\uparrow}$ and spin-down $\psi_{\downarrow}$ particles
             $$ i\hbar\partial_t\psi_{\uparrow}=\biggl((-i\hbar\nabla-\frac{q}{c}\textbf{A})^2+q\varphi-\gamma B_z\biggr)\psi_{\uparrow}$$\begin{equation}\label{H1}\qquad\qquad\qquad\qquad\qquad\qquad\qquad-\gamma(B_x-iB_y)\psi_{\downarrow}
                       \end{equation}
                      {\em and}

                        $$ i\hbar\partial_t\psi_{\downarrow}=\biggl((-i\hbar\nabla-\frac{q}{c}\textbf{A})^2+q\varphi+\gamma B_z\biggr)\psi_{\downarrow}$$\begin{equation}\label{H2}\qquad\qquad\qquad\qquad\qquad\qquad\qquad-\gamma(B_x+iB_y)\psi_{\uparrow}.
                       \end{equation}
where $\varphi$ and $\textbf{A}$  - are scalar and vector potentials, respectively, $B_{\alpha}$ - are the components of magnetic field. The first step in the        construction of QHD apparatus is
to determine the spin current of particles in the neighborhood
of $\textbf{r}$ in a physical space.  In the consideration of spin-up particles and spin-down particles as two different species the spin current takes the form of

$$\textbf{J}_x(\textbf{r},t)=\frac{1}{2m}\Biggr(\psi^{+}_{\downarrow}\textbf{D}^{\beta}\psi_{\uparrow}+
\textbf{D}^{+}\psi_{\downarrow}^{+}\psi_{\uparrow}$$

\begin{equation}  \label{JM1}\qquad\qquad\qquad\qquad+\psi^{+}_{\uparrow}\textbf{D}^{\beta}\psi_{\downarrow}+
\textbf{D}^{+}\psi_{\uparrow}^{+}\psi_{\downarrow}\Biggl),
\end{equation}{\em  and}

$$\textbf{J}_y(\textbf{r},t)=i\frac{1}{2m}\Biggr(\psi^{+}_{\downarrow}\textbf{D}^{\beta}\psi_{\uparrow}+
\textbf{D}^{+}\psi_{\downarrow}^{+}\psi_{\uparrow}$$

\begin{equation}  \label{JM2}\qquad\qquad\qquad\qquad-\psi^{+}_{\uparrow}\textbf{D}^{\beta}\psi_{\downarrow}-
\textbf{D}^{+}\psi_{\uparrow}^{+}\psi_{\downarrow}\Biggl).
\end{equation}

where $\hat{D}^{\alpha}=-i\hbar\hat{\nabla}^{\alpha}-q/cA^{\alpha}$.

Differentiation of spin current definitions (\ref{JM1}) and (\ref{JM2}) with respect to time and applying
of  Schrodinger equations  (\ref{H1}) and (\ref{H2})   we obtain the spin current equations
for charged spinning particles     in the form of

$$\partial_{t}\textbf{J}_{x}+\partial_{\beta}J_{x}^{\alpha\beta}=\frac{2\gamma}{\hbar}B_z\textbf{J}_y-\frac{2\gamma}{\hbar}B_y\textbf{J}_z$$

\begin{equation}
\label{Jx}+\frac{q}{mc}\textbf{J}_x\times\textbf{B}+\frac{q}{m}S_x\textbf{E}+\frac{\gamma}{m}(n_{\uparrow}+n_{\downarrow})\nabla B_x,
\end{equation}
{\em and}

$$\partial_{t}\textbf{J}_y+\partial_{\beta}J_{y}^{\alpha\beta}=-\frac{2\gamma}{\hbar}B_z\textbf{J}_x+\frac{2\gamma}{\hbar}B_x\textbf{J}_z$$

\begin{equation} \label{Jy}+\frac{q}{mc}\textbf{J}_y\times\textbf{B}+\frac{q}{m}S_y\textbf{E}+\frac{\gamma}{m}(n_{\uparrow}+n_{\downarrow})\nabla B_y.
\end{equation}

Equations (\ref{Jx}) and (\ref{Jy}) are  derived for the charged spinning particles.  The flux of the spin-current $J_{h}^{\alpha\beta}$  which emerges in the second term at the  left-hand side of equations (\ref{Jx}) and (\ref{Jy}) contains physical quantities defined for spin-up and spin-down particles. It was found that the flux of the spin-current has complex structure and contains additional contribution of quantum kinematics as the quantum Bohm potential

$$J_{x}^{\alpha\beta}=\frac{1}{2}J_x^{\alpha}(v^{\beta}_{\uparrow}+v^{\beta}_{\downarrow})-\frac{\hbar}{4m}S_y\nabla^{\alpha}(v^{\beta}_{\uparrow}-v^{\beta}_{\downarrow})$$

\begin{equation} \label{JJx}-\frac{\hbar}{8m}S_y(v^{\alpha}_{\uparrow}+v^{\alpha}_{\downarrow})\biggl(\frac{\nabla^{\beta}n_{\uparrow}}{n_{\uparrow}}
-\frac{\nabla^{\beta}n_{\downarrow}}{n_{\downarrow}} \biggr)+S_x\Delta^{\alpha\beta},
\end{equation}
{\em and}

$$J_{y}^{\alpha\beta}=\frac{1}{2}J_y^{\alpha}(v^{\beta}_{\uparrow}+v^{\beta}_{\downarrow})+\frac{\hbar}{4m}S_x\nabla^{\alpha}(v^{\beta}_{\uparrow}-v^{\beta}_{\downarrow})$$

\begin{equation} \label{JJy}+\frac{\hbar}{8m}S_y(v^{\alpha}_{\uparrow}+v^{\alpha}_{\downarrow})\biggl(\frac{\nabla^{\beta}n_{\uparrow}}{n_{\uparrow}}
-\frac{\nabla^{\beta}n_{\downarrow}}{n_{\downarrow}}\biggr)-S_y\Delta^{\alpha\beta},\end{equation}
where the quantum potential contribution $\Delta^{\alpha\beta}$ takes the form of

   $$  \Delta^{\alpha\beta}=-\frac{\hbar^2}{4m^2}\biggl( \frac{\partial^{\alpha}\partial^{\beta}\sqrt{n_{\uparrow}}}{n_{\uparrow}}+ \frac{\partial^{\alpha}\partial^{\beta}\sqrt{n_{\downarrow}}}{n_{\downarrow}}\biggr)$$

   \begin{equation} \qquad\qquad+
     \frac{\hbar^2}{4m^2}\biggl( \frac{\partial^{\alpha}\sqrt{n_{\uparrow}}\partial^{\beta}\sqrt{n_{\downarrow}}}{n_{\uparrow}n_{\downarrow}}+ \frac{\partial^{\alpha}n_{\downarrow}\partial^{\beta}n_{\uparrow}}{n_{\uparrow}n_{\downarrow}}\biggr)           \end{equation}

The first term in the definitions (\ref{JJx}) and (\ref{JJy}) describes the flux of spin current generated by the spin motion in the system of spin-up and spin-down particles, the second  terms are neglected when $\textbf{v}_{\uparrow}=\textbf{v}_{\downarrow}$. The third and fourth terms represent the quantum Bohm potential influence in the case of different population of states with different spin direction.  We do not consider the contribution of thermal motion to the spin current evolution.

The electric $\textbf{E}$  and magnetic $\textbf{B}$ fields appearing in
the equations (\ref{j1}) - (\ref{Jy}) include the contribution of the classical charge current and the spin density sources and   satisfy to
the Maxwell equations, such as {\em Ampere's law}  where  the total  current
produced by the free charges  and includes the magnetization spin current of charged particles ($q_e$ and  $q_i$)

$$
\nabla\times \textbf{B}=\frac{4\pi}{c}(q_en_{\uparrow}\textbf{v}_{\uparrow}+q_en_{\downarrow}\textbf{v}_{\downarrow}+q_in_i\textbf{v}_i) $$

\begin{equation}\label{Amper}\qquad\qquad\qquad\qquad\qquad\qquad+
4\pi\gamma\nabla\times\textbf{S}+\frac{1}{c}\frac{\partial\textbf{E}}{\partial t}, \end{equation}
and {\em Feraday's law}

      \begin{equation}    \nabla\times \textbf{E}=- \frac{1}{c}\frac{\partial\textbf{B}}{\partial t}  \end{equation}
where the magnetic moment density and magnetization current are defined as $\textbf{M}=(\gamma S_x, \gamma S_y, \gamma(n_{\uparrow}-\rho_{\downarrow}))$  and $\Im^{\alpha}_{\beta}=\{\gamma\textbf{J}_x, \gamma\textbf{J}_y, \gamma(n_{\uparrow}\textbf{v}_{\uparrow}-n_{\downarrow}\textbf{v}_{\downarrow})\}.$

\section{Perturbation evolution}
  We study collective eigen-waves in a system of neutral
spinning particles considering spin-up and spin-down states as two different species and being in an external uniform magnetic
field. We have deal with the paramagnetic and
diamagnetic dielectrics. We have used the  self-consistent field approximation and derived the set of spin-current equations for degenerate particles. But the real paramagnetic and
diamagnetic dielectrics  must be described in the context of nondegenerate  systems.  In this paper we are focused on the basic  new dispersion properties of waves and do not take into account the effects of nondegenerate particles.

  We use equations of quantum hydrodynamics, where physical quantities are determined in terms of contributions from both spin-up and spin-down particles, to study dispersion of the collective excitations in the three dimensional samples of the magnetized dielectrics. In this paper we apply a two-fluid model of spin-up and spin-down states to study wave processes in magnetized dielectrics.  We take into account the case when the equilibrium spin current $\textbf{J}_{\alpha0}$ equals to zero.

  Let's consider propagation of waves in magnetized dielectric making distinction between spin-up and spin-down states. To do that we can analyze small monochromatic perturbations of physical variables from the stationary state

\begin{equation}\label{di} n_{\uparrow,\downarrow} =n_{0\uparrow,\downarrow}+\delta n_{\uparrow,\downarrow},  \qquad  \textbf{v}_{\uparrow,\downarrow}=0+\delta \textbf{v}_{\uparrow,\downarrow}, \end{equation}
 $$\textbf{B}=\textbf{z}B_{0}+\delta \textbf{B}, \qquad \textbf{E}=0+\delta \textbf{E}, $$ $$ S_{x}=\delta S_x, \qquad S_{y}=\delta S_y, \qquad \textbf{J}_{\alpha}=0+\delta\textbf{J}_{\alpha}. $$

We can mark, in this section and below, all physical quantity is presented in the form of sum of equilibrium part and small perturbations
$$f=f_{0}+\delta f.$$

In this case if we assume that linear excitations $\delta f$ are
proportional to $exp(-\imath\omega t+\imath\textbf{k}\textbf{r})$
a linearized set of equations (\ref{n1}), (\ref{n2}), (\ref{j1}), (\ref{j2}) and (\ref{s1}), (\ref{s2}), (\ref{JJx}), (\ref{JJy}).

     \subsection{Perpendicular propagation}

We choose the external magnetic field as $B_0 = \textbf{z}B_{0}$ with respect to the wave propagation direction determined by the wavenumber $k = k_y$ of the wave.  We use our formalism to study the oblique propagation of  the waves in the system of neutral spinning particles. The first order electromagnetic field are $\delta\textbf{E}=E\textbf{x}$ and $\delta\textbf{B}=B\textbf{z}=-ckE/\omega\textbf{z}.$   We taken into account  the low-frequency
 wave which propagates perpendicular
to the background magnetic field   and the energy flux density moves along the propagation
direction.  The low-frequency mode is obtained when $\omega^2 << k^2c^2$. After some calculations we obtain the dispersion equation
\begin{equation}\label{w1}
1+\frac{\Omega_{s\uparrow}^2}{\omega^2-v^2_{\uparrow}k^2}+\frac{\Omega_{s\downarrow}^2}{\omega^2-v^2_{\downarrow}k^2}=0,
\end{equation}
where we use the definition  for the thermal velocity of each species $v^2_{\uparrow,\downarrow}=2^{2/3}v^2_{F\uparrow,\downarrow}/3+\hbar^2k^2/4m^2$ and $v_{F\uparrow,\downarrow}=(3\pi^2n_{0\uparrow,\downarrow})^{1/3}\hbar/m$ - is the Fermi velocity. The evolution of spin-current leads to existence of faster damping spin current waves \cite{100} with frequency which can be generalized for the case of separately described spin-up  and spin-down states $$\Omega_{s\uparrow,\downarrow}^2=-\frac{4\pi\gamma^2n_{0\uparrow,\downarrow}}{m}k^2$$.

The dispersion equation (\ref{w1}) has the {\em general solution}
      \begin{widetext}
\begin{equation}\label{w2}\omega^2_{\pm}=\frac{1}{2}\biggl(v^2_{\uparrow}k^2+v^2_{\downarrow}k^2-\Omega_{s\uparrow}^2-\Omega_{s\downarrow}^2
\pm
\sqrt{(v^2_{\uparrow}k^2-v^2_{\downarrow}k^2)^2+(\Omega_{s\uparrow}^2+\Omega_{s\downarrow}^2)^2+2(v^2_{\uparrow}k^2-v^2_{\downarrow}k^2)(\Omega_{s\downarrow}-\Omega_{s\uparrow})} \biggr)     \end{equation}      \end{widetext}

           {\em Assuming that the external magnetic field is weak} and, as a result $n_{0\uparrow,\downarrow}=n_0/2\mp\triangle n/2,$ where the difference between concentrations of spin-up and spin down of particles is $\triangle n=n_{0\uparrow}-n_{0\downarrow}$. In this approximation and in the limit of small difference between concentrations of spin-up and spin-down particles $\triangle n$  the solution (\ref{w2}) takes the form

 \begin{widetext}   \begin{equation}\label{w3}\omega^2_{+}=\frac{1}{3}v^2_{F}k^2\biggl(1-\frac{1}{9}(\frac{\triangle n}{n_0})^2\biggr)+
    \frac{\hbar^2}{4m^2}k^4+\delta,\end{equation}

 {\em   and}

 \begin{equation} \label{w4} \omega^2_{-}=-\frac{4\pi\gamma^2k^2n_{0}}{m}+\frac{1}{3}v^2_{F}k^2\biggl(1-\frac{1}{9}(\frac{\triangle n}{n_0})^2\biggr)
  +
    \frac{\hbar^2}{4m^2}k^4-\delta,\end{equation}    \end{widetext}
                                {\em   where}

       \begin{equation}\label{w5}
       \delta=\frac{1}{9}v^2_{F}(\frac{\triangle n}{n_0})^2\frac{m}{4\pi\gamma^2n_0}\times
       \biggl(\frac{4}{9}v^2_{F}k^2+\frac{8\pi\gamma^2k^2n_{0}}{m}\biggr)\end{equation}

  The first solution (\ref{w3}) describes a {\em sound-like wave} existing in magnetized dielectrics due to different equilibrium distribution of spin-up and spin-down particles, where $v_{F}=(3\pi^2n_{0})^{1/3}\hbar/m$ - is the total Fermi velocity. The quantum pressure produces a nonlinear contribution $\sim k^4$  that may be important at large $k$. The second solution (\ref{w4}) represents the {\em spin-current wave} dispersion in the system of spin-up and spin-down particles, where $n_{0}=n_{0\uparrow}+n_{0\downarrow}$.  The quantum correction associated to the Bohm potential is taken into account. For small wave numbers and in the case when  the total concentration is $n_0>\pi\hbar^6/m^3\gamma^6$, the dispersion solution (\ref{w4}) shows faster damping and possesses no wave behavior. But for large $k$  wave number and for $n_0=10^{22}$cm$^{-3}$, $\triangle n/n_0\approx 10^{-2}$ the quantum correction $\sim\hbar^2$ produces a nonlinear contribution and modifies the spectrum of spin current wave.

                     In strong external magnetic fields $B_0\geq 10^4 G$ all spins must be polarized antiparallel to the field  $n_0\approx n_{0\downarrow}$. But we can use the small amount of spin-up particles $n_{0\uparrow}=\xi$ and $n_{0\downarrow}=n_0-\xi.$ In this case the dispersion solution has the limit
            \begin{equation}\label{w6}\omega^2_{+}=\frac{2^{2/3}}{3}v^2_{F}k^2\biggl(1-\frac{2}{3}\frac{\xi}{n_0}\biggr)+
    \frac{\hbar^2}{4m^2}k^4,\end{equation}
         {\em  and}

  \begin{equation}\label{w7}\omega^2_{-}=-\frac{4\pi\gamma^2k^2n_{0}}{m}+\frac{2^{2/3}}{3}v^2_{F}k^2\biggl(1-\frac{2}{3}\frac{\xi}{n_0}\biggr)
  +
    \frac{\hbar^2}{4m^2}k^4.\end{equation}

Lets discuss the limit of small spin wave frequency $ 4\pi\gamma^2n_{0}/m << v_{F}^2.$  This approximation can be relevant for the magnetized dielectrics with  the density $n_0 < 10^{21}$ cm$^{-3}$.  In this approximation the      solution (\ref{w2}) takes the form

      $$ \omega^2_{+}=\frac{1}{3}v^2_{F\uparrow}k^2\biggl(1-(\frac{\triangle n}{n_0})\biggr)+
    \frac{\hbar^2}{4m^2}k^4$$\begin{equation}\label{w8}\qquad\qquad\qquad\qquad-\frac{2\pi\gamma^2k^2n_{0}}{m}-\lambda,\end{equation}
                       {\em  and}

  $$\omega^2_{-}=\frac{1}{3}v^2_{F\downarrow}k^2\biggl(1-(\frac{\triangle n}{n_0})\biggr)+
    \frac{\hbar^2}{4m^2}k^4$$\begin{equation}\label{w9}\qquad\qquad\qquad\qquad-\frac{2\pi\gamma^2k^2n_{0}}{m}+\lambda.\end{equation}

 $$ \lambda=\frac{4\pi\gamma^2n_{0}}{m}k^2\times\biggl(  \frac{1+\frac{8}{9} v^2_{Fe}k^2(\frac{\triangle n}{n_0})^2}{\frac{16}{9}v^2_{F}\frac{\triangle n}{n_0}}  \biggr)
 $$
 where the solutions (\ref{w8}) and (\ref{w9})  separately describe spin-up  and spin-down  sound-like waves.

  \section{Conclusion}
The interests and publications of research in spin dynamics and transports till increased. There are many theoretical publications that propose new definitions of spin current \cite{24} - \cite{26}. But we have used the  quantum hydrodynamics (QHD) description  which separately describes spin-up   and spin-down particles. We start with the Pauli equation  for a single particle \cite{12}, \cite{133}, \cite{23}.

To make more detailed study of magnetic moment dynamics the equation of spin-current evolution  for the unpolarized systems was
derived in Ref. \cite{100} and new type of collective excitations in magnetized dielectrics
                               which were called spin-current waves was predicted.
  We have derived the spin current evolution equations for charged spin-1/2 particles  (\ref{JJx}) and (\ref{JJy}) considering dynamics of particles with spin-up and spin-down separately. We predicted the contribution of quantum corrections due to the Bohm potential and difference between spin-up and spin-down particle velocities to the dynamics of spin current.

  In Ref. \cite{12}  authors  studied propagation of waves parallel to external magnetic field without spin-current effects and   found contribution of magnetic field in the Langmuir wave dispersion via difference of occupation of spin-up and spin-down states.    The  oblique propagation of longitudinal waves without spin-current excitations had been studied in Ref. \cite{23} and new solutions besides   of two well-known waves (the Langmuirwave and the Trivelpiece–Gould wave) had been found.
  Implication of these spin-current evolution equations (\ref{JJx}), (\ref{JJy}) leads to novel wave solutions. In this paper we have studied the propagation of spin-acoustic wave in magnetized dielectrics with spin-polarized particles under the action of an external magnetic field on the magnetic moments. We have analyzed the limits of weak and strong magnetic fields and determined the influence of spin polarization on the wave solutions.
  Assuming that the external magnetic field is weak we found the sound-like wave existing in magnetized dielectrics (\ref{w3}).  A sound-like wave appears to arise in systems of neutral particles due to different equilibrium distribution of spin-up and spin-down states.

  It was found that spin-current evolution in magnetized dielectrics leads to the formation of new types of waves $\omega=\pm i\sqrt{4\pi n_0/m}\gamma k$  \cite{100}.
One of them has increasing amplitude while another has decreasing amplitude. We have derived the spin-current wave in  the system of spin-polarized particles and have found the quantum contribution to the spin current wave (\ref{w4}).
It was shown that the  dependence of dispersion on $\triangle n/n_0$ is quadratic at small magnetization. These solution disappears if the negative term $\sim \Omega^2_s$ caused by the spin current dominates over the Fermi pressure.

We have found the solution in strong external magnetic fields (\ref{w6}) and (\ref{w7}).
 The dispersion solution for spin-current wave in strong magnetic field (\ref{w7}) shows linear dependence of $\omega^2_{-}$ on  $\triangle n/n_0$.

Method of the QHD can become very  powerfull method of studying influence of magnetic
moment dynamics on various processes in magnetic nanostructures.
  Presented here results can be important for the processes of spin transport  in ferromagnetic quantum wires,  for the theoretical investigation of optical spin injection, spin flip in
the result of electron interaction with electromagnetic wave,  for the theoretical investigation of spin-diode structures, spin-transistor and another spintronic devices.  It allows to investigate wave processes  and   introduce new phenomenon in linear and non-linear  regimes of small amplitude perturbations, for example to find the soliton solution for the spin-polarized current in a  ferromagnetic nanowire.  The more detailed consideration of spin transport phenomena   will be considered during further development and application of the spin separated QHD model with spin-current description  developed in this paper.  For this purpose we need to describe  the spin transport  in
presence of spin-orbit interaction.

  \bibliographystyle{elsarticle-num}

\end{document}